\documentclass[12pt]{cernart}

\newcommand{\gevc }{GeV/$c$}
\newcommand{\DG }{$\left< {\Delta g}/{g} \right>_{x}$~}
\newcommand{\dg }{\left< \frac{\Delta g}{g} \right>_{x}}
\usepackage{graphicx}

\begin {document}
\dimen\footins=\textheight

\begin{titlepage}
\docnum{CERN--PH--EP/2008--003}
\date{8th February 2008}
\vspace{1cm}

\title{\LARGE Direct Measurement of the Gluon Polarisation in
       the Nucleon via Charmed Meson Production}

\vspace*{0.5cm}
\collaboration{The COMPASS Collaboration}

\vspace{2cm}
\begin{abstract}
We present the first measurement of the gluon polarisation in the nucleon
based on the photon--gluon fusion process tagged by charmed meson
production and decay to charged $K$ and $\pi$. The data were collected
in polarised muon scattering off a polarised deuteron target
by the COMPASS
collaboration at CERN during 2002--2004. The result of this LO analysis
is $\dg=-0.47 \pm 0.44 (\mbox{stat}) \pm 0.15 (\mbox{syst})$ at
$\langle x_{\rm} \rangle
\approx 0.11$ and a scale $\mu^2\approx 13\,(\mbox{GeV}/c)^2$.
\end{abstract}

\vspace*{60pt}
\vfill
\submitted{submitted to Physical Review Letters}

\begin{Authlist}
{\large  The COMPASS Collaboration}\\[\baselineskip]
{\small
%
%
M.~Alekseev\Iref{turin_p},
V.Yu.~Alexakhin\Iref{dubna},
Yu.~Alexandrov\Iref{moscowlpi},
G.D.~Alexeev\Iref{dubna},
A.~Amoroso\Iref{turin_u},
A.~Arbuzov\Iref{dubna},
B.~Bade\l ek\Iref{warsaw},
F.~Balestra\Iref{turin_u},
J.~Ball\Iref{saclay},
J.~Barth\Iref{bonnpi}
% ...
}
\end{Authlist}

%
%
\Instfoot{bielefeld}{Universit\"at Bielefeld, Fakult\"at f\"ur Physik, 33501 Bielefeld, Germany\Aref{e}}
\Instfoot{bochum}{Universit\"at Bochum, Institut f\"ur Experimentalphysik, 44780 Bochum, Germany\Aref{e}}
\Instfoot{bonniskp}{Universit\"at Bonn, Helmholtz-Institut f\"ur  Strahlen- und Kernphysik, 53115 Bonn, Germany\Aref{e}}
\Instfoot{bonnpi}{Universit\"at Bonn, Physikalisches Institut, 53115 Bonn, Germany\Aref{e}}
\Instfoot{brno}{Institute of Scientific Instruments, AS CR, 61264 Brno, Czech Republic\Aref{f}}
\Instfoot{burdwan}{Burdwan University, Burdwan 713104, India\Aref{g}}
\Instfoot{calcutta}{Matrivani Institute of Experimental Research \& Education, Calcutta-700 030, India\Aref{h}}
\Instfoot{dubna}{Joint Institute for Nuclear Research, 141980 Dubna, Moscow region, Russia}
\Instfoot{erlangen}{Universit\"at Erlangen--N\"urnberg, Physikalisches Institut, 91054 Erlangen, Germany\Aref{e}}
\Instfoot{freiburg}{Universit\"at Freiburg, Physikalisches Institut, 79104 Freiburg, Germany\Aref{e}}
\Instfoot{cern}{CERN, 1211 Geneva 23, Switzerland}
\Instfoot{heidelberg}{Universit\"at Heidelberg, Physikalisches Institut,  69120 Heidelberg, Germany\Aref{e}}
\Instfoot{helsinki}{Helsinki University of Technology, Low Temperature Laboratory, 02015 HUT, Finland  and University of Helsinki, Helsinki Institute of  Physics, 00014 Helsinki, Finland}
\Instfoot{liberec}{Technical University in Liberec, 46117 Liberec, Czech Republic\Aref{f}}
\Instfoot{lisbon}{LIP, 1000-149 Lisbon, Portugal\Aref{i}}
\Instfoot{mainz}{Universit\"at Mainz, Institut f\"ur Kernphysik, 55099 Mainz, Germany\Aref{e}}
\Instfoot{miyazaki}{University of Miyazaki, Miyazaki 889-2192, Japan\Aref{j}}
\Instfoot{moscowlpi}{Lebedev Physical Institute, 119991 Moscow, Russia}
\Instfoot{munichlmu}{Ludwig-Maximilians-Universit\"at M\"unchen, Department f\"ur Physik, 80799 Munich, Germany\AAref{e}{k}}
\Instfoot{munichtu}{Technische Universit\"at M\"unchen, Physik Department, 85748 Garching, Germany\AAref{e}{k}}
\Instfoot{nagoya}{Nagoya University, 464 Nagoya, Japan\Aref{j}}
\Instfoot{praguecu}{Charles University, Faculty of Mathematics and Physics, 18000 Prague, Czech Republic\Aref{f}}
\Instfoot{praguectu}{Czech Technical University in Prague, 16636 Prague, Czech Republic\Aref{f}}
\Instfoot{protvino}{State Research Center of the Russian Federation, Institute for High Energy Physics, 142281 Protvino, Russia}
\Instfoot{saclay}{CEA DAPNIA/SPhN Saclay, 91191 Gif-sur-Yvette, France}
\Instfoot{telaviv}{Tel Aviv University, School of Physics and Astronomy, 69978 Tel Aviv, Israel\Aref{l}}
\Instfoot{triest_i}{Trieste Section of INFN, 34127 Trieste, Italy}
\Instfoot{triest}{University of Trieste, Department of Physics and Trieste Section of INFN, 34127 Trieste, Italy}
\Instfoot{triestictp}{Abdus Salam ICTP and Trieste Section of INFN, 34127 Trieste, Italy}
\Instfoot{turin_u}{University of Turin, Department of Physics and Torino Section of INFN, 10125 Turin, Italy}
\Instfoot{turin_i}{Torino Section of INFN, 10125 Turin, Italy}
\Instfoot{turin_p}{University of Eastern Piedmont, 1500 Alessandria,  and Torino Section of INFN, 10125 Turin, Italy}
\Instfoot{warsaw}{So{\l}tan Institute for Nuclear Studies and Warsaw University, 00-681 Warsaw, Poland\Aref{m} }
\Instfoot{warsawtu}{Warsaw University of Technology, Institute of Radioelectronics, 00-665 Warsaw, Poland\Aref{n} }
\Instfoot{yamagata}{Yamagata University, Yamagata, 992-8510 Japan\Aref{j} }
%
%
\Anotfoot{+}{Deceased}
\Anotfoot{a}{Also at IST, Universidade T\'ecnica de Lisboa, Lisbon, Portugal}
\Anotfoot{b}{On leave of absence from JINR Dubna}
\Anotfoot{c}{Also at Chubu University, Kasugai, Aichi, 487-8501 Japan}
\Anotfoot{d}{Also at GSI mbH, Planckstr.\ 1, D-64291 Darmstadt, Germany}
\Anotfoot{e}{Supported by the German Bundesministerium f\"ur Bildung und Forschung}
\Anotfoot{f}{Suppported by Czech Republic MEYS grants ME492 and LA242}
\Anotfoot{g}{Supported by DST-FIST II grants, Govt. of India}
\Anotfoot{h}{Supported by  the Shailabala Biswas Education Trust}
\Anotfoot{i}{Supported by the Portuguese FCT - Funda\c{c}\~ao para a Ci\^encia e Tecnologia grants POCTI/FNU/49501/2002 and POCTI/FNU/50192/2003}
\Anotfoot{j}{Supported by the Ministry of Education, Culture, Sports, Science and Technology, Japan, Grant-in-Aid for Specially Promoted Research No.\ 18002006; Daikou Foundation and Yamada Foundation}
\Anotfoot{k}{Supported by the DFG cluster of excellence `Origin and Structure of the Universe' (www.universe-cluster.de)}
\Anotfoot{l}{Supported by the Israel Science Foundation, founded by the Israel Academy of Sciences and Humanities}
\Anotfoot{m}{Supported by Ministry of Science and Higher Education grant 41/N-CERN/2007/0 and the MNII research funds for 2005-2007}
\Anotfoot{n}{Supported by KBN grant nr 134/E-365/SPUB-M/CERN/P-03/DZ299/2000}
\end{titlepage}

\thispagestyle{empty}
\setcounter{page}{0}

Pioneering experiments on the spin structure of the nucleon
performed in the seventies \cite{vernon} were
followed by the EMC experiment at CERN which obtained
a stunning conclusion on the quark contribution to the proton spin \cite{emc}.
This result triggered
extensive studies of the spin structure of the nucleon 
in lepton scattering experiments at CERN by
the SMC \cite{smc} and COMPASS \cite{compass} 
SLAC \cite{e155_d}, DESY (HERMES) \cite{hermes} and JLAB
\cite{jlab} as well as in polarised proton-proton collisions at RHIC \cite{phenix,star}. 
The results of these
studies confirmed the validity of the Bjorken sum rule and 
the violation of the Ellis--Jaffe sum rule. In addition, 
the parton helicity distributions in the nucleon were extracted using
QCD analyses. The quark contribution
to the proton helicity is now confirmed to be around 0.3, smaller than 
the expected value of 0.6 \cite{refa8}. However, due to the limited 
range in $Q^2$ covered by the experiments at fixed $x_{\rm Bj}$ the
QCD analyses
(e.g.\cite{compass}) show limited sensitivity to the gluon helicity
distribution, $\Delta g(x)$, and its first moment, $\Delta G$.
The determination of  $\Delta g(x)$ has therefore to be
complemented by direct measurements in dedicated experiments.  

The gluon polarisation \DG
has been determined from the
photon--gluon fusion (PGF) process
by HERMES \cite{hermes_highpt}, SMC \cite{SMC_highpt} and COMPASS
\cite{compass_highpt_lowq}. These analyses used events containing hadron pairs
with high transverse momenta, $p_{\rm T}$, with respect to the virtual photon
direction.  
This method provides good statistical precision but relies
on Monte Carlo generators simulating QCD
processes. The measurements point towards a small
value of the gluon polarisation at $x\approx 0.1$. This is in line
with recent results from PHENIX \cite{phenix}
and STAR \cite{star} at RHIC.

In the Quark Parton Model the nucleon spin is given by the quark spins,
$\Delta \Sigma$, while $\Delta G$ vanishes. Taking into account
orbital angular momenta, $L$, of quarks and gluons the nucleon spin
is 
\begin{equation}
\frac{1}{2} = \frac{1}{2} \Delta \Sigma + \Delta G + L_z \,.
\end{equation}
In QCD the U(1) anomaly generates
a  gluonic contribution to the measured singlet axial coupling,
$a_0(Q^2)$. This anomalous gluonic contribution does not vanish at 
$Q^2\rightarrow \infty$. As a result, $\Delta\Sigma (Q^2)$ becomes
scheme dependent and may differ from the observable $a_0$
while $\Delta G$ is scheme--independent at least up to the NLO. 
In the Adler--Bardeen factorization scheme \cite{ABscheme}
$\Delta\Sigma^{\rm AB}$ is independent of $Q^2$.
Restoring the Ellis-Jaffe value of $\Delta\Sigma^{\rm AB}\approx 0.6$ requires 
a value of $\Delta G(Q^2)\approx 2$ and $L_z \approx -2$ at $Q^2 = 5 $~(GeV/$c)^2$.

Here, we present a new result on \DG from muon-deuteron
scattering. 
The gluon polarisation is determined assuming that open charm 
production is dominated by the
PGF mechanism yielding a $c\bar{c}$ pair which fragments
   mainly into $D$ mesons.
This method has the advantage that in lowest order there are no other
contributions to the cross section; however, it is statistically
limited.
In our analysis only one
charmed meson is required in every event. This meson  is
selected through its decay in one of the two channels:
$D^{*}(2010)^+ \rightarrow D^0\pi^+_{\rm slow}\rightarrow K^-\pi^+\pi^+_{\rm
  slow}$  ($D^*$ sample) and $D^0\rightarrow K^-\pi^+$  ($D^0$
sample) and their charge conjugates.

The data
were collected during 2002 to 2004 with the 160~GeV/$c$ CERN SPS $\mu^+$ beam
and correspond to an integrated luminosity of 1.7~fb$^{-1}$. The
beam muons coming from the $\pi^+$ and $K^+$ decays are
naturally polarised with an average polarisation,
$P_{\rm b}$, of about $-80$\%.

The polarised $^6$LiD target consists of two cells (upstream $u$,
downstream $d$), each 60~cm long, longitudinally polarised with 
opposite orientations.
The spin directions are reversed every eight hours by rotating the
field of the target magnet system. 
The average target polarisations, $P_{\rm t}$, were $\pm 50$\%.
The $^6$Li nucleus basically consists of an $^4$He core plus a
deuteron. A
dilution factor, $f$, of about 0.4 is obtained for the target material.
The exact value is kinematics dependent and is calculated as described
in \cite{compass_A1}.
Particle tracking  and identification are performed in a two-stage
spectro\-meter \cite{spectrometer}.

The present analysis selects events with an incoming muon, a scattered
muon identified behind hadron absorbers, an interaction
vertex
in the target and at least two additional charged tracks.
The $D^0$ mesons are reconstructed through their $K\pi$ decay
which has a branching ratio of 3.8\%. Due to
multiple Coulomb scattering of the charged particles in the 
solid state target the spatial resolution of the vertex reconstruction
is not sufficient to separate the $D^0$ production and
decay vertices. Therefore, the mesons are reconstructed on a combinatorial
basis, considering all pairs of oppositely charged tracks
in a given event and
calculating their invariant mass. This method results in a high combinatorial 
background which is reduced in further analysis steps.

The largest background reduction stems from kaon identification
in the Ring Imaging CHerenkov counter (RICH).
This restricts the studied events to a sample with
kaons of momenta exceeding 9.1~\gevc.
Simulations have shown that in the acceptance about
70\% of kaons coming from
$D^0$ decays exceed this threshold.

Particle identification in the RICH starts from reconstructed
tracks with measured momenta. The angles
between the track and the detected Cherenkov photons are calculated
for each track. 
The comparison of the expected angular distribution of photons  
for a pion, a kaon or a proton with the measured one is used for particle
identification. For this comparison two different methods were used.
The data from 2002 and 2003 were analysed using a
$\chi^2$ calculated from the photon angles and the Cherenkov angles.
The mass hypothesis with the smallest $\chi^2$ is selected.
For the 2004 data,
the likelihoods for the various mass hypotheses are also compared to
the background likelihood. The background likelihood function
is evaluated using photons not
associated to reconstructed tracks.

\begin{figure}[tb]
\begin{center}
\includegraphics[width=0.8\hsize,clip]{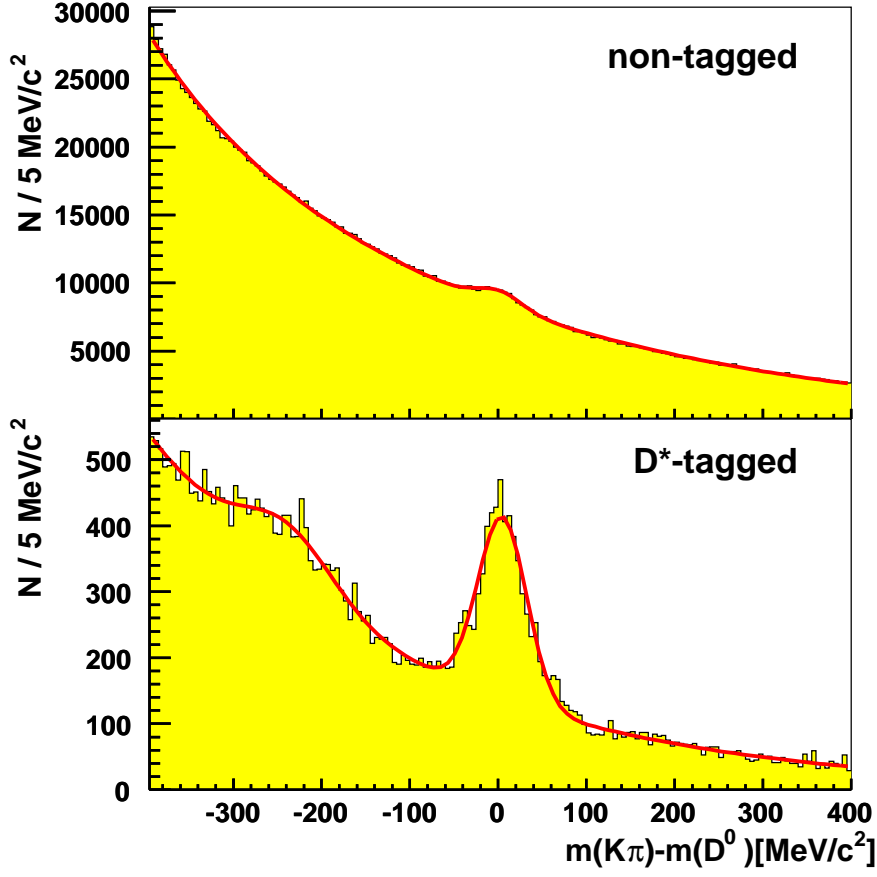}
\end{center}
\caption{\small Invariant mass distributions of the $K\pi$ pairs tagged with the
$D^*$ decay ($D^*$ sample, lower) and of the non-tagged $K\pi$ pairs
($D^0$ sample, upper). The
curves represent fits with functional form described in the text.}
\label{d0fit_all}
\end{figure}

In the analysis  an identified kaon  and an identified pion are
required for each event except for the $D^*$ sample
analysed with the  $\chi^2$ method where all tracks not identified as kaons
are considered as pion candidates.
The $D^*$ and the $D^0$ samples are analysed independently.
The following two kinematic cuts are used for the $D^*$ ($D^0$) sample:
a cut $z>0.2$ ($z>0.25$), where $z$ is the fraction of the energy 
of the virtual photon carried by the $D^0$ meson candidate, and a cut
$|\mathrm{cos}\theta^*|<0.85$ ($|\mathrm{cos}\theta^*|<0.5$),
where $\theta^*$ is the decay angle in the $D^{0}$ c.m.s. system.
In the $D^*$ channel a cut on the
mass difference is imposed, $3.1 \mathrm{\ MeV/}c^2 < M_{K\pi\pi_{slow}} -
M_{K\pi} - M_{\pi}< 9.1 \mathrm{\ MeV/}c^2$, where $M_{K\pi\pi_{slow}}$
and $M_{K\pi}$ are the masses of the $D^*$ and the $D^{0}$
candidates, respectively.
The resulting signal-to-background ratio  for events in the signal
region is approximately 1:1 for the
$D^*$ and 1:10 for the $D^0$ sample
(Fig.~\ref{d0fit_all}). 
Note that the events entering the $D^*$ sample
are not used in the $D^0$ sample.

The number of events, $N$, collected in a given target cell
and time interval (about one week of data taking) is 
{\small 
\begin{eqnarray}\label{N}
  N = a \phi n (\sigma_{\rm S} + \sigma_{\rm B}) \times~~~~~~~~~~~~~~~~~~~~~~~~~~~ \nonumber \\ 
  \left [1 + P_{\rm t} P_{\rm b} f \left(a_{\rm LL}
          \frac{\sigma_{\rm S}}{\sigma_{\rm S} + \sigma_{\rm B}} \dg + D
         \frac{\sigma_{\rm B}}{\sigma_{\rm S} + \sigma_{\rm B}}
  A_{\rm B} \right)\right ]\,\,.
\end{eqnarray}
}
Here, $a$, $\phi$ and $n$
are the spectrometer acceptance, the integrated incident muon flux and
the number of target nucleons, respectively. 
In addition $\sigma_{\rm S}$ ($\sigma_{\rm B}$) is the cross section
of the events described by the central Gaussians (background)
in Fig.~1.
The analysing power $a_{\rm LL}$ is the asymmetry
for the $\mu g \rightarrow \mu c\bar{c}$ process, and $A_{\rm B}$ the background 
asymmetry. The depolarisation factor, $D$, from Ref.~\cite{compass_lowq} is used.

A simultaneous extraction of \DG and $A_{\rm B}$, assumed to be
constant in the mass range considered,
is performed independently for the $D^*$ and the $D^0$ sample using
the events in the mass range $-400$~MeV/$c^2
< M_{D^0} - M_{K\pi} < 400$~MeV/$c^2$ 
recorded in the two target cells before ($u$,$d$) and
after ($u'$,$d'$)
target spin reversal. The events of the four samples are
weighted with a signal weight, $w_{\rm S}$, and independently with a
background weight, $ w_{\rm B}$,
\begin{equation}
  w_{\rm S} = P_{\rm b} f a_{\rm LL} \frac{\sigma_{\rm S}}{\sigma_{\rm
  S} + \sigma_{\rm B}}\,\,,\,\,
    w_{\rm B} = P_{\rm b} f D \frac{\sigma_{\rm B}}{\sigma_{\rm S} +
    \sigma_{\rm B}}\,\,.
\end{equation}
The target polarisation is not included into the weights because it is
time dependent. In this way 8 equations 
\begin{equation}\label{W}
\sum_{i=1}^{N_t} w^t_{{\rm C},i}=\alpha^t_{\rm C}\left(1+
\beta^t_{\rm C} \dg +\gamma^t_{\rm C} A_{\rm B}\right ) 
\end{equation}
with 
 \begin{eqnarray}
\beta^t_{\rm C}
=\frac{\sum_i^{N_t} P_{{\rm t},i} w^t_{{\rm S},i} w^t_{{\rm C},i}}{\sum_i^{N_t} w^t_{{\rm C},i}}\,\,,\,\,
\gamma^t_{\rm C}
=\frac{\sum_i^{N_t} P_{{\rm t},i} w^t_{{\rm B},i} w^t_{{\rm C},i}}{\sum_i^{N_t}
w^t_{{\rm C},i}} 
\end{eqnarray}
are obtained from Eq.~(\ref{N}) for the 10 unknowns which are
\DG, $A_{\rm B}$
and 8 acceptance factors $ \alpha_{\rm C}^{t}=\int a^t \phi^t n^t (\sigma_{\rm
  S} + \sigma_{\rm B})  w^t_{\rm C} {\rm d}X$ 
with $t=u,d,u',d'$ and ${\rm C=S,B}$ \cite{pretz}. 
Here, $\int {\rm d}X$ stands for the integration over the accessible kinematic region.
Assuming that possible acceptance variations affect the upstream and
downstream cells in the same way, i.e. 
   ${\alpha_{\rm C}^{u}/ \alpha_{\rm C}^{d}}
    =    {\alpha_{\rm C}^{u'}/ \alpha_{\rm C}^{d'}}$,
provides two additional equations (as in Refs.~\cite{compass,compass_lowq}).
With an extra, much weaker, assumption that signal and background 
events on the same target cell
are affected in the same way by the acceptance variations
one finally arrives at a system of 8 equations with 7 unknowns.
Possible deviations from the above assumptions generate false
asymmetries which are included in the systematic 
error. 

For the evaluation of Eq.~(\ref{W}) $P_{\rm b}$ is parameterized as
a function of the beam momentum. 
For $P_{\rm t}$,  values averaged over about one hour are used.
The signal purity ${\sigma_{\rm S}}/{(\sigma_{\rm S}+\sigma_{\rm
B})}$, is obtained from a
fit to the $M_{K\pi}-M_{D^0}$ spectra.
The fit is done separately for the events originating from the two
target cells as well as for five (three) separate bins of
$f P_{\rm b} a_{\rm LL}$ for the $D^*$ ($D^0$) sample. 
This takes care of the correlation between the signal
purity and the analysing power. 
The spectra are fitted by the sum of signal and background
functions, described by a Gaussian and  a product of an exponential
and a polynomial, respectively. In
case of the $D^{*}$ sample a second Gaussian is used 
to describe the reflection of $D^{\,0}\,\rightarrow\,
K\,\pi\,\pi^{\,0}$ decay, where the $\pi^{\,0}$ meson is
not observed. In the $D^0$ sample this reflection
is not visible. The total number
of $D^0$ mesons  is about 3,800 and 13,800
in the $D^*$ and the $D^0$ samples, respectively. 
\begin{figure}[tbp]
\begin{center}
\includegraphics[width=0.8\hsize,clip]{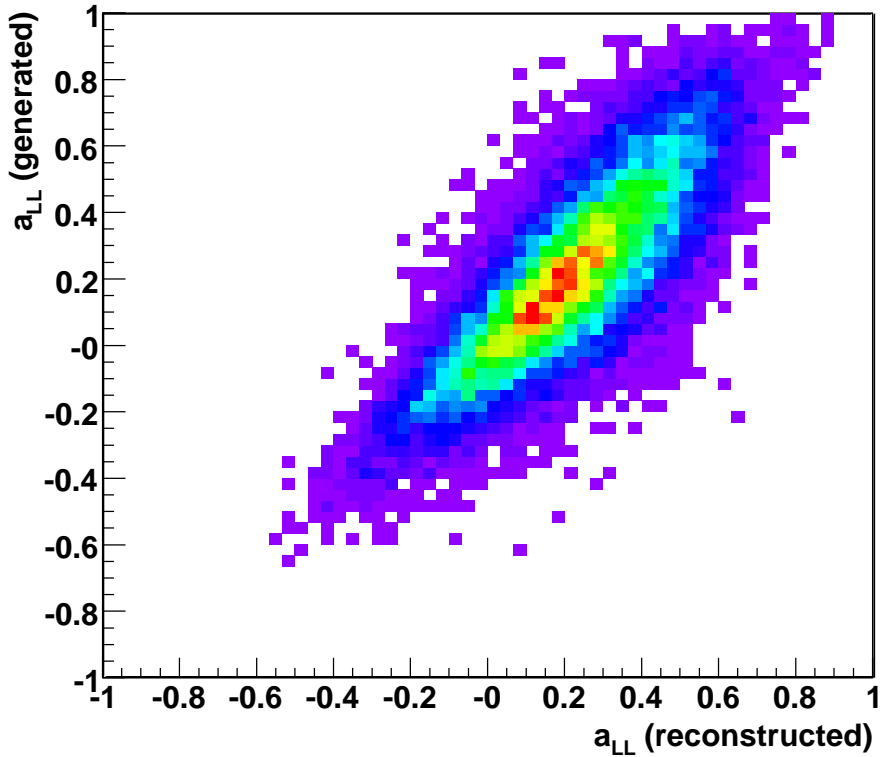}
\end{center}
\caption{\small Correlation between generated and reconstructed values
of $a_{\rm LL}$.}
\label{a_ll}
\end{figure}

Finally, the determination of \DG requires 
know\-ledge of the analysing power $a_{\rm LL}$. 
Since only one $D^0$ is measured, the partonic 
kinematics cannot be fully reconstructed and 
$a_{\rm LL}$ cannot be calculated on an event-by-event basis.
A kinematic factor which is approximately equal to the depolarisation
factor $D$ is factored out and the remaining part of $a_{\rm LL}$
is parameterized in terms of measured
kinematic variables.
This is done using  a neural
network trained on a Monte Carlo (MC) sample for $D^*$ mesons.
The correlation between the generated
and  the reconstructed $a_{\rm LL}$ is 82\% (see Fig.~\ref{a_ll}).
The sample was generated
with AROMA \cite{aroma} in leading order QCD and the events were 
processed by GEANT to simulate the response of
the detector and finally reconstructed like real events.
The scale, $\mu^2$, used in the MC simulation was determined by the  
mass of the produced charm quark pair and is sufficiently large
to justify the perturbative approach.

The gluon polarisation \DG is determined for each of the 29 weeks of
data taking with
a standard least square minimisation procedure taking into account
the statistical correlation between events weighted by $w_{\rm S}$ and
by $w_{\rm B}$. 
As in our inclusive analysis \cite{compass} the event weighting reduces 
the statistical error.
The final value for  \DG is the weighted mean of the above results.
The resulting $A_{\rm B}$ is consistent with zero.
Note that we measure $\Delta g/g$ in a given range of $x$.
Provided that ${\Delta g}/{g(x)}$
is weakly dependent on $x$ in the range covered, 
this method gives a measurement of $\Delta g/g(\langle x \rangle)$, where $\langle x \rangle$ is calculated with
signal weights. The above assumption is supported by the results of
our QCD analysis \cite{compass}.

\begin{table}[tp]
\caption{\small Systematic error contributions to \DG.}
\label{tab:D0}
\begin{center}
\begin{tabular}{lc||lc}
\hline
\hline
source & $\delta (\frac{\Delta g}{g})$ & source &$\delta 
(\frac{\Delta g}{g})$ \\
\hline
\hline
False asymmetry   & $ 0.09$&Beam polarization $P_{\rm b}$   &  0.02  \\
Fitting   & $0.09$&Target polarization $P_{\rm t}$   & 0.02\\
Binning   & $0.04$&Dilution factor $f$ & 0.02 \\
MC parameters   & $0.05$ &&\\
\hline
\multicolumn{4}{c}{Total error~~~~0.15}\\
\hline
\hline
\end{tabular}
\end{center}
\end{table}

\begin{figure}[tbp]
\begin{center}
\includegraphics[width=0.8\hsize,clip]{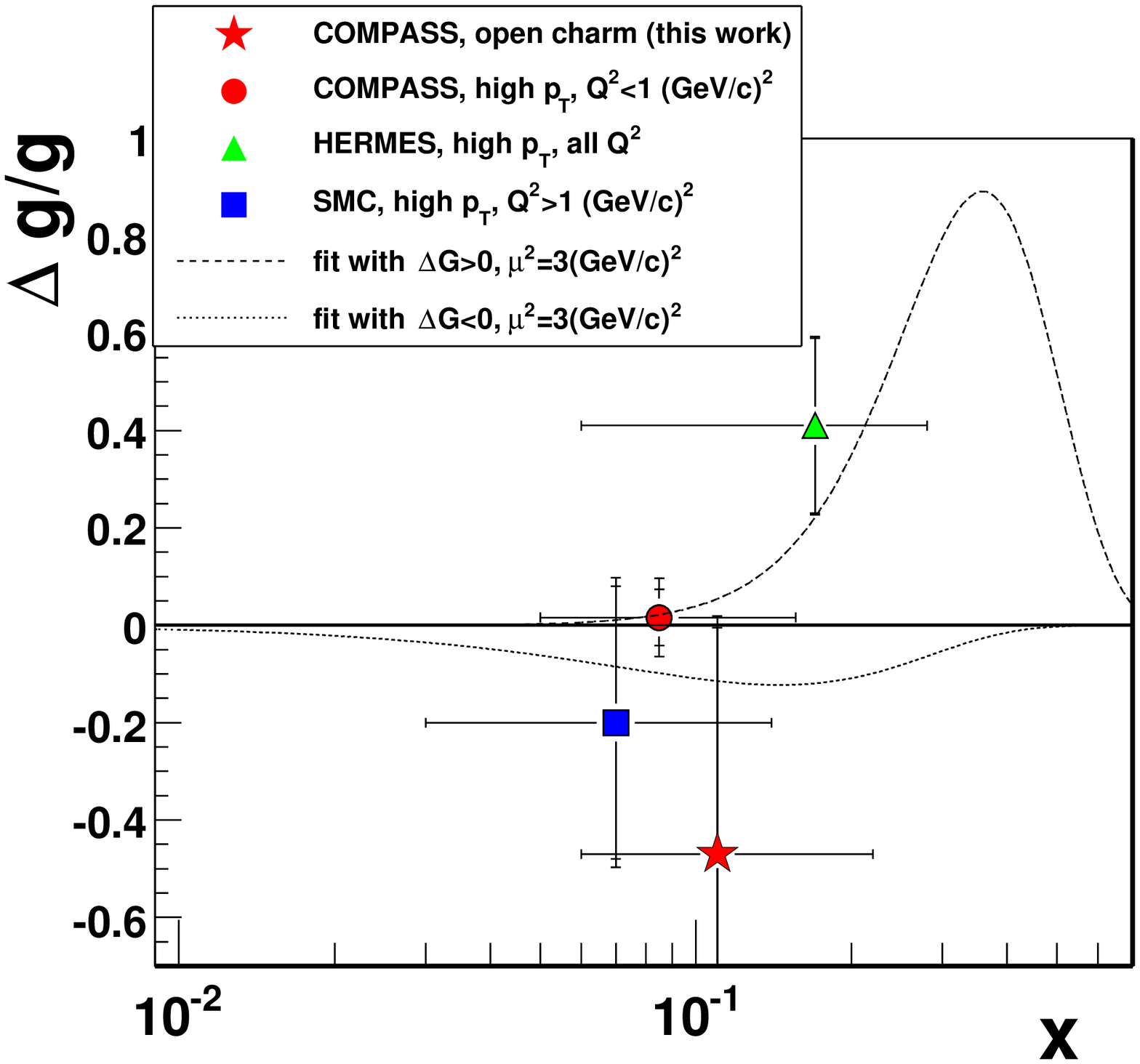}
\end{center}
\caption{\small Compilation of the \DG measurements from
open charm and high $p_{\rm T}$ hadron pair production by COMPASS 
\cite{compass_highpt_lowq},
SMC \cite{SMC_highpt} and HERMES \cite{hermes_highpt} as a function of $x$.
Horizontal bars mark the range in $x$ for each measurement, vertical
ones give the statistical precision and the total errors (if available). 
The open charm measurement is at
a scale of about 13~(GeV/$c$)$^2$, other measurements at 3~(GeV/$c$)$^2$.
The curves display  parameterizations from a 
NLO QCD analysis in the $\overline{\rm MS}$ scheme 
at 3~(GeV/$c$)$^2$, \cite{compass}: fits with $\Delta G > 0$ (broken line)
and with $\Delta G < 0$ (dotted line).}
\label{DGG_all}
\end{figure}
The major contributions to the systematic error are listed in 
Table~\ref{tab:D0}. The contributions from $P_{\rm b}$, $P_{\rm t}$ and $f$ 
are as discussed in \cite{compass}. 
To study the influence of false asymmetries the final samples from
Fig.~1 were subdivided 
into two samples using criteria related to the experimental apparatus, 
e.g.~kaons going to the upper or to the lower spectrometer parts. 
The resulting asymmetries were found to be compatible
within their statistical accuracy, thus no false asymmetries were observed.
The upper limit of the contribution to the systematic error was also
estimated from the dispersion
of the values for \DG and $A_B$ for the various data taking week.
Assuming that possible detector 
instabilities are similar for background and signal events and applying a 
method as in [4] leads to a conservative limit of $0.09$ for both
decay channels.
The fit for the signal purity determination was performed
with different background parameterizations, different
binnings for the invariant mass spectra and varying the constraints
for 
some of the fit parameters.
The resulting dispersion of the \DG values was 0.09 for both channels.
The \DG
calculations were repeated with several sets of binning in
$fP_{b}a_{\rm LL}$
and  the dispersion of the results was 0.04 for both channels.
Other contributions like radiative
corrections and event migration between target cells
are negligible. 
All these studies were done independently for the $D^0$ and
$D^*$ samples and result in very similar values for all the contributions.
To estimate the influence of the simulation parameters,
i.e.\ charm quark mass (1.3~GeV/$c^2$ to 1.6~GeV/$c^2$)
and parton distribution functions, MC samples with different parameter
sets were generated and $a_{\rm LL}$ was recalculated.
The dispersion of the resulting \DG for the combined $D^0$ and $D^*$
sample was 0.05. In addition, it was
checked
that the parameterization of $a_{\rm LL}$ is valid for the $D^0$ and
the $D^*$ sample. 
The resolved photon contribution to the open charm production via
gluon-gluon fusion has been estimated with the RAPGAP generator
\cite{rapgap} and found to be negligible in our kinematic range. 

The values obtained for \DG  are $0.53\pm 0.75 (\mbox{stat})$ for the $D^0$ and
$-1.01\pm 0.55 (\mbox{stat})$ for the $D^*$ sample. These two values agree within $1.7$
standard deviations and their weighted mean is 
 $$\dg =-0.47\pm 0.44(\mbox{stat})\pm 0.15(\mbox{syst})$$at a value of $\langle x \rangle = 0.11^{+0.11}_{-0.05}$ and a scale $\langle
\mu^2\rangle \approx 13~$(GeV/$c$)$^2$. All the contributions to the
systematic error in Table~\ref{tab:D0} were added in quadrature and
conservatively assumed to be fully correlated in the two samples.

In Fig.~\ref{DGG_all} the above result is compared
to other measurements of \DG and to parameterizations from a QCD analysis
of the structure function data \cite{compass}. 
It is in good agreement with previous measurements
favouring small values of \DG. Note that the scale here is
$\mu^2\approx 13$~(GeV/$c$)$^2$ while all other points and the curves are given
at $\mu^2\approx 3$~(GeV/$c$)$^2$.

In summary, we have performed the first determination of \DG 
from a measurement of the cross section asymmetry for $D^0$ meson
production.
In the analysis photon--gluon fusion in LO QCD was assumed to be the
underlying production mechanism for open charm production.
The resulting value of
\DG is compatible with our previous result from high $p_{\rm T}$
hadron pairs but is less model dependent.

We acknowledge the support of the CERN management and staff, the
special
effort of CEA/Saclay for the target magnet project, as well as
the skills and efforts of the technicians of the collaborating
institutes. 



%

\end{document}